\documentclass[conference]{IEEEtran}
\IEEEoverridecommandlockouts
\usepackage{cite}
\usepackage{amsmath,amssymb,amsfonts}
\usepackage{algorithmic}
\usepackage{algorithm}
\usepackage{booktabs}
\usepackage{graphicx}
\usepackage{textcomp}
\usepackage{xcolor}
\usepackage{bm}
\def\BibTeX{{\rm B\kern-.05em{\sc i\kern-.025em b}\kern-.08em
    T\kern-.1667em\lower.7ex\hbox{E}\kern-.125emX}}
\begin{document}

\title{LLM4MAC: An LLM-Driven Reinforcement Learning Framework for MAC Protocol Emergence}

\author{Renxuan Tan, Rongpeng Li, and Zhifeng Zhao
\thanks{R. Tan and R. Li are with the College of Information Science and Electronic Engineering, Zhejiang University (\{ttrx,lirongpeng\}@zju.edu.cn).
Z. Zhao is with Zhejiang Lab as well as Zhejiang University (zhaozf@zhejianglab.com).
}

}
\newcommand\subscriptuet[1][n]{^{\text{u}_{#1}}_t}
\newcommand\subscriptbst{^{\text{b}}_t}
\newcommand\subscriptbsuet[2][t]{^{\text{b},\text{u}_{#2}}_{#1}}
\newcommand\UEn[1][n]{\text{u}_{#1}}
\newcommand\BS{\text{BS}}

\newcommand\ueact[1][n]{a^{\text{u}_{#1}}_{t,p}}
\newcommand\uemsg[1][n]{a^{\text{u}_{#1}}_{t,s}}

\newcommand{\uepolicy}{\pi^{{\rm u}_n}}
\newcommand{\bspolicy}{\pi^{{\rm b}_n}}
\newcommand{\llmpolicy}{\pi^{\rm llm}}
\maketitle

\begin{abstract}
With the advent of 6G systems, emerging hyper-connected ecosystems necessitate agile and adaptive medium access control (MAC) protocols to contend with network dynamics and diverse service requirements. We propose LLM4MAC, a novel framework that harnesses large language models (LLMs) within a reinforcement learning paradigm to drive MAC protocol emergence. By reformulating uplink data transmission scheduling as a semantics-generalized partially observable Markov game (POMG), LLM4MAC encodes network operations in natural language, while proximal policy optimization (PPO) ensures continuous alignment with the evolving network dynamics. A structured identity embedding (SIE) mechanism further enables robust coordination among heterogeneous agents. Extensive simulations demonstrate that on top of a compact LLM, which is purposefully selected to balance performance with resource efficiency, the protocol emerging from LLM4MAC outperforms comparative baselines in throughput and generalization.
\end{abstract}

\begin{IEEEkeywords}
Protocol  Emergence, Large Language Models, Reinforcement Learning, Medium Access Control
\end{IEEEkeywords}

\section{Introduction}
\begin{figure*}[t]
    \centering
    \includegraphics[width = 0.95\linewidth]{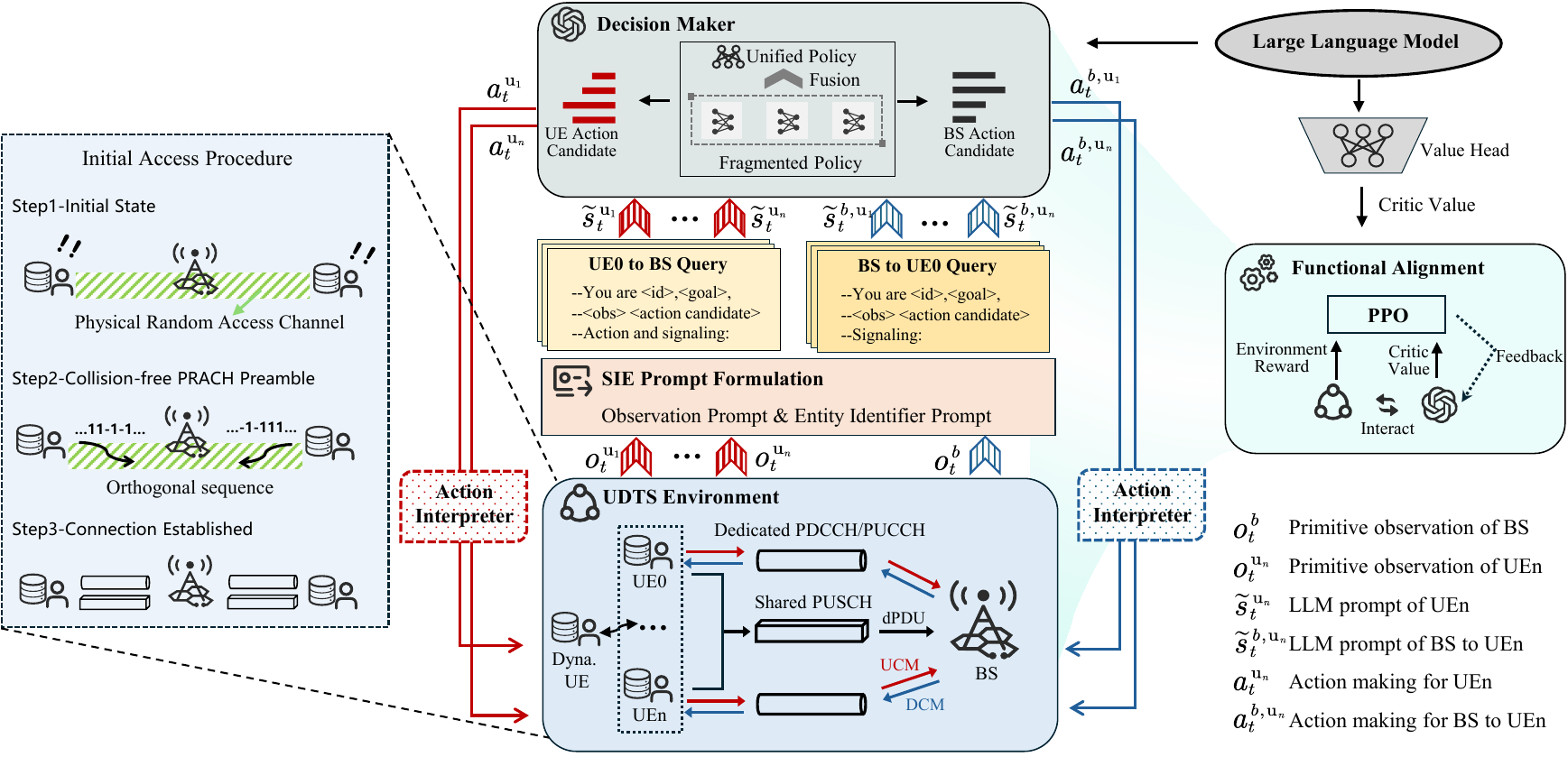}
    \caption{LLM4MAC framework consists of three main components: prompt formulation transforms observations into identifiable model inputs, decision maker determines actions, and environmental feedback with PPO for functional alignment.}
    \vspace{-1em}
    \label{fig:framework}
\end{figure*}
The vision for intelligent, hyper-connected, and data-intensive ecosystems \cite{IMT2030} poses unprecedented challenges for wireless medium access control (MAC) protocol design, where human-crafted protocols struggle to 
deliver bespoke quality-of-service (QoS) guarantees. Recently, multi-agent reinforcement learning (MARL) has emerged as a potent paradigm by fostering emergent communication among distributed agents \cite{miuccioLearningGeneralizedWireless2022}. Towards optimizing uplink data transmission scheduling (UDTS) in scenarios encompassing a fixed number of user equipment (UE), prior studies 
\cite{MP-scale} have modeled UEs and base stations (BS) as cooperative agents and attained marked improvements in access policy learning and signaling efficiency. Similarly, RL-based approaches have been leveraged to dynamically orchestrate protocol features, yielding adaptive designs that outperform IEEE 802.11ac in throughput and latency\cite{keshtiarastWirelessMACProtocol2024}. However, the inherent rigidity of fixed agent numbers during training often renders the resulting policies brittle in the face of stochastic network dynamics, due to intermittent arrivals and departures of agents.

In parallel, large language models (LLMs) have demonstrated exceptional capabilities in comprehension, generalization, and reasoning, positioning them as promising tools for resolving intricate protocol and communication challenges \cite{wang2024nextgenerationwifinetworksgenerative,shrestha2024adaptinglargelanguagemodels,Wu2024NetLLMAL,chen_netgpt_2024}. Prevailing approaches typically rely on prompt or in-context learning with meticulously curated supervised datasets to yield context-aware solutions for TCP optimization \cite{shrestha2024adaptinglargelanguagemodels}, network traffic prediction \cite{Wu2024NetLLMAL}, and Wi-Fi protocol design \cite{wang2024nextgenerationwifinetworksgenerative}. Nonetheless, such methodologies are often constrained by dataset-specific path dependencies. 
In other words, compared to symbolic MARL \cite{MP-scale,miuccioLearningGeneralizedWireless2022,keshtiarastWirelessMACProtocol2024}, 
LLM-oriented prompt learning approaches provide superior generalization capability, but 
are often limited in protocol emergence scenarios where optimal interaction behaviors must be discovered through exploration. 

In light of these challenges, we introduce LLM4MAC — a robust protocol learning framework that reconceptualizes dynamic UDTS environments as LLM-reinforced interactions across agents, thereby transcending the performance ceiling of individual LLM and RL approaches. 
Particularly, we reformulate the MAC layer scheduling task as a semantics-generalized extension of partially observable Markov games (POMGs), where network operations — including signaling messages, actions, and observations — are calibrated in natural language. 
To tackle the underlying multi-agent coordination task, the LLM, deployed at the BS, simultaneously orchestrates both BS and UE operations in a contextual manner, while the orchestration leverages a hierarchical, structured identity embedding (SIE) mechanism to enrich contextual differentiation and alignment. A unified policy fusion mechanism is proposed to merge fragmented agent policies into a coherent framework. Moreover, by harnessing Proximal Policy Optimization (PPO) \cite{carta2024groundinglargelanguagemodels} for continuous, feedback-driven adaptation, our framework achieves resilient functional alignment with evolving network dynamics. Extensive simulations substantiate that LLM4MAC not only navigates the inherent complexities of dynamic UDTS but also delivers significant enhancements over traditional MARL and LLM prompt learning approaches, which require extensive post-processing to map context-infused responses \cite{carta2024groundinglargelanguagemodels}, accordingly paving the way for emerging more adaptive, efficient, and generalizable MAC protocols.

This work is structured as follows. Section II presents the system model and formulates the problem. Section III details the LLM-based decision maker and its PPO training. Section IV evaluates performance via simulations and the main conclusion is drawn in Section V.
\section{System Model and Problem Formulation}
\subsection{System Model }
\subsubsection{Background on UDTS}
As illustrated in Fig. \ref{fig:framework}, a BS serves $N$ UEs, denoted by $\mathcal{N} = \{\UEn[1], \cdots, \UEn[N]\}$, where some dynamic UEs may arrive or departure. Consistent with 3GPP TS 38.321 \cite{3GPP38321}, we assume that the BS has assigned each UE dedicated uplink/downlink control channels (PUCCH/PDCCH) and a contention-based shared channel (PUSCH). Notably, the channel allocation between the BS and UEs must adopt an efficient MAC protocol that jointly optimizes signaling and access strategies, since concurrent transmissions on the PUSCH can trigger collisions that impede the BS’s ability to decode dPDUs. For pre-defined protocols \cite{MP-new}, at time $t$, UE $\UEn$ exchange uplink control messages (UCMs) - with specific meanings like \texttt{scheduling requests}, \texttt{buffer reports}, or \texttt{NACKs} - with the BS, which in turn issues downlink control messages (DCMs) such as \texttt{grants}, \texttt{ACKs}. 
These signaling exchanges, generally conveyed over the dedicated PUCCH and PDCCH, are critical for coordinating access and setting the operational context for data delivery. Afterward, UEs deliver MAC data PDUs (dPDUs) to the BS via the PUSCH under a time division multiple access (TDMA) scheme, with a maximum episode length $T$.
However, such pre-defined protocols often exhibit rigidity and redundancy. 
For example, under ideal channels, some \texttt{ACKs} can be saved to boost throughput \cite{MP-0}. Correspondingly, learnable, context-aware messages can yield superior results \cite{Semantic-paper}.

\subsubsection{Delivery Model}
Without loss of generality, each UE transmits a learnable UCM $\uemsg \in \{1,\cdots,U\}$ and the BS reciprocally feeds back a DCM $a\subscriptbsuet{n} \in \{1,\cdots,D\}$ to $\UEn$, where $U$ and $D$ denote vocabulary sizes. Assuming a Poisson process as in  \cite{MP-scale}, UE $\UEn$ generates a dPDU with probability $p_a$ per time slot, denoted by $\xi^n_{t,\text{gen}} \in \{0,1\}$. 
Meanwhile, each UE is equipped with a first-in-first-out (FIFO) buffer of capacity $\lvert \mathcal{B}\rvert$ to temporarily store dPDUs, with the buffer occupancy denoted by $b\subscriptuet \in \{0, 1, \dots, \lvert \mathcal{B} \rvert\}$.
Moreover, a UE can determine a transmission action $\ueact \in \{0,1,2\}$, corresponding to \texttt{do nothing}, \texttt{transmit dPDU}, or \texttt{delete dPDU}, respectively. Excellent MAC protocols shall maximize BS's dPDU reception during $T$ and minimize UE's buffer occupancy, where $\xi_{t,\text{rec}} \in \{0,1\}$ is the received dPDU at BS. More importantly, these protocols shall competently adapt to variations in UE numbers.

\subsubsection{Channel Model}
The PUSCH is modeled as a packet erasure channel, subject to a non-zero transport block error rate (TBLER) \cite{miuccioLearningGeneralizedWireless2022}. The instantaneous state of the data channel is represented by $x_t \in \{0,\cdots,N+1\}$, where $0$ denotes idle, $N+1$ signifies a collision (i.e., no decodable energy), and $x_t=n\in \{1, \dots, N\}$ specifies that the $n$-th UE is transmitting. The BS acquires channel state information (CSI) via the demodulation reference signal (DMRS) \cite{3GPP38321}, enabling informed coordination and adaptive control of the MAC protocol.

\subsubsection{Decision Model}
Based on the observations, a decision-making model governs both UE-BS signaling and the UE's transmission behavior. 
For a UE $\UEn$, define its observation at time $t+1$ as $o^{{\rm u}_n}_{t+1}=(b\subscriptuet,a\subscriptuet,a\subscriptbsuet{n})$ whereas the taken action  $a\subscriptuet=(\ueact,\uemsg)$. Meanwhile, the BS's observation is given by $o^{\rm b}_{t+1}=(x_t,\uemsg[0],a\subscriptbsuet{0},\cdots,\uemsg[N],a\subscriptbsuet{N})$ and its action is defined as $a\subscriptbst=(a\subscriptbsuet{1},\cdots,a\subscriptbsuet{N})$. Crucially, both BS and UE observations are primarily composed of preceding interactive UCMs/DCMs, which evolve into meaningful signaling messages during the training process and thereby dynamically refine UE behaviors to optimize throughput.
\subsection{Semantic-Generalized POMG Formulation}
In the UDTS environment, the learning of MAC protocols — where agents make decisions (i.e., signaling
messages and buffering actions) based on individual observations and coordinate to avoid collisions — can be modeled as a POMG. Due to the much easier maintenance of semantic consistency, we extend the POMG framework by incorporating semantics through a dimension-aware linguistic representation. 
As mentioned lately, despite the variations in UE numbers, this method contributes to avoiding inconsistent representation learning.
Specifically, we introduce a language vocabulary $\mathcal{W}$ and a mapping function set $\bm f$. The BS's linguistic observation space $\Tilde{\mathcal{O}}^{\BS} \subset \mathcal{W}$ is then defined as 
\begin{equation}
    \Tilde{\mathcal{O}}^{\BS} = \{ \Tilde{o}\subscriptbst | \Tilde{o}\subscriptbst = f_{\mathcal{O}^\BS}(o\subscriptbst),\forall o\subscriptbst\in \mathcal{O}^\BS \},\ f_{\mathcal{O}^\BS}\in {\bm f},
\end{equation}
where $f_{\mathcal{O}^\BS}$ transforms numerical observations into linguistic representations. Notably, for simplicity of representations, we use a tilde notation to differentiate the linguistic observation space $\Tilde{\mathcal{O}}$ (resp. linguistic action space $\Tilde{\mathcal{A}}$) with primitive observation space $\mathcal{O}$ (resp. action space $\mathcal{A}$). By adopting identical mapping operations to observations and actions of UEs, we can have linguistic observation space $\Tilde{\mathcal{O}}^{\UEn}$ and action space $\Tilde{\mathcal{A}}^{\UEn}$ for agent $\UEn$, as well as linguistic action space $\Tilde{\mathcal{A}}^{\BS}$ for BS. We denote the joint linguistic  observation space $\Tilde{\mathcal{O}}:=\Tilde{\mathcal{O}}^{\UEn[1]} \times \cdots \times \Tilde{\mathcal{O}}^{\BS}$ and the joint linguistic action space of all agents $\Tilde{\mathcal{A}}:=\Tilde{\mathcal{A}}^{\UEn[1]} \times \cdots \times \Tilde{\mathcal{A}}^{\BS}$. Therefore, the semantic-generalized POMG can be attained as 
$(\Tilde{\mathcal{O}},\Tilde{\mathcal{A}},\mathcal{T},r,\mathcal{W},{\bm f},T,\gamma)$,
where $\mathcal{T}: \Tilde{\mathcal{O}} \times \Tilde{\mathcal{A}} \mapsto \Tilde{\mathcal{O}}$ is the transition function and the reward function $r:\Tilde{\mathcal{O}} \times \Tilde{\mathcal{A}} \mapsto \mathbb{R}$ can be given as
\begin{equation}
\label{eq:reward}
r_t= 
\begin{cases}
+\varrho, & \text {if new dPDU received }(\xi_{t,{\rm rec}} =1); \\
-\varrho, & \text {if unreceived dPDU deleted}; \\
0, & \text {otherwise},
\end{cases}
\end{equation}
where $\varrho$ is a positive constant that rewards successful dPDU reception at the BS and penalizes the erroneous deletions of unreceived dPDUs. Each episode in the UDTS environment spans $T$ time slots, with the cumulative return defined as
\begin{equation}
\label{eq:cum_reward}
    G = \sum\nolimits_{t=0}^{T-1} \gamma^t r_t,
\end{equation}
where $\gamma\in (0,1]$ is the discount factor. To maximize system throughput and reduce collisions, implicitly accomplished by maximizing Eq. \eqref{eq:cum_reward}, we resort to the integration of LLM and RL within the semantic-generalized POMG framework for joint optimization of transmission actions and signaling.
\begin{figure*}[t]
    \begin{equation}
    \label{eq:critic_loss}
    L(\phi)=\mathbb{E}_{\tau}\left[ \sum\nolimits_{t=1}^T\max \left[\left(V(s_t)-\hat{R}_t\right)^2,\left(\operatorname{clip}\left(V(s_t), V_{\text {old }}(s_t)-\epsilon, V_{\text {old }}(s_t)+\epsilon\right)-\hat{R}_t\right)^2\right]\right]
    \end{equation}
\hrulefill
\end{figure*}
\section{PPO-based LLM Decision Maker for Protocol Emergence}
Building on the semantic-generalized POMG formulation, this section presents a framework where a single LLM functions as the centralized decision-maker for all involved agents. PPO grounds LLM performance through functional alignment with the dynamic environment, while SIE-augmented prompts facilitating agent differentiation. As illustrated in Fig. \ref{fig:framework}, LLM4MAC encompasses three key components, namely the SIE prompt formulation mechanism, an LLM decision maker $\llmpolicy$, and the function alignment strategy via PPO.
\subsection{Preliminaries on PPO}
We employ PPO, an efficient and robust RL algorithm, to optimize the LLM-based decision maker (i.e., actor) $\llmpolicy(a|s)$ and a critic network $V(s)$ parameterized by $\theta$ and $\phi$, where $a$ and $s$ represent action and state, respectively\footnote{For simplicity of representation, we slightly abuse the notations in this part.}. The training objective tries to update the actor and critic networks, thus locating an optimal policy that maximizes Eq. \eqref{eq:cum_reward}. Updates to the critic network involve minimizing squared temporal-difference residuals as in Eq. \eqref{eq:critic_loss}, given on the top of Page \pageref{eq:critic_loss}, where $\hat{R}_t = \sum_{t^\prime = t}^{T} \gamma^{t^{\prime}-t} r_{t^\prime}$ is the discounted reward-to-go. A clipping function $\text{clip}(\cdot)$ limits updates relative to old value estimate $V_{\rm old}$. The trajectory
$\tau$ in Eq. \eqref{eq:rollout} is defined as 
\begin{equation}
    \label{eq:rollout}
     \tau = \left\{ \left(s_t,a_t,r_t,\llmpolicy(a_t \lvert s_t),V(s_t),A_t
    \right)\right\}_{t=0}^{\lvert T-1\rvert}
    ,
\end{equation}
captures state, action, reward, policy, current value, and the generalized advantage $A_t$ that is computed via $A_t = \sum_{l=0}^{T}(\gamma\lambda)^l(r_{t+l} + \gamma V(s_{t+l+1}) - V(s_{t+l}))$ with discount factor $\lambda=0.9$.
The actor-network parameters are updated using the sampled policy gradient of Eq. \eqref{eq:actor_loss} on Page \pageref{eq:critic_loss}, in which $\eta_{\theta,t} =  \frac{\llmpolicy(a_t \lvert s_t)}{\pi_{\rm old}^{\rm llm}(a_t \lvert s_t)}$ denotes the important weight, the operator $D_\text{KL}(\cdot\|\cdot)$ computes the KL-divergence, the policy entropy $H$ encourages exploration, and $\epsilon$ is a hyperparameter to limit magnitude of updates. 

\begin{figure*}[t]
    \begin{equation}
    \label{eq:actor_loss}
    J(\theta) = \mathbb{E}_{\tau}\left[
     \sum\nolimits_{t=1}^T  \left[ \min \left(\eta_{\theta,t} A_t, \operatorname{clip}\left(\eta_{\theta,t}, 1-\epsilon, 1+\epsilon\right) A_{t}\right) - D_\text{KL}(\llmpolicy_{\rm old}(\cdot|s_t)\lvert\lvert\llmpolicy(\cdot|s_t))  + H(\llmpolicy)\right]\right]
\end{equation}
    \hrulefill
\end{figure*}
\subsection{LLM-Driven Protocol Emergence with PPO Alignment}
\begin{figure}[t]
    \centering
    \includegraphics[width = \linewidth]{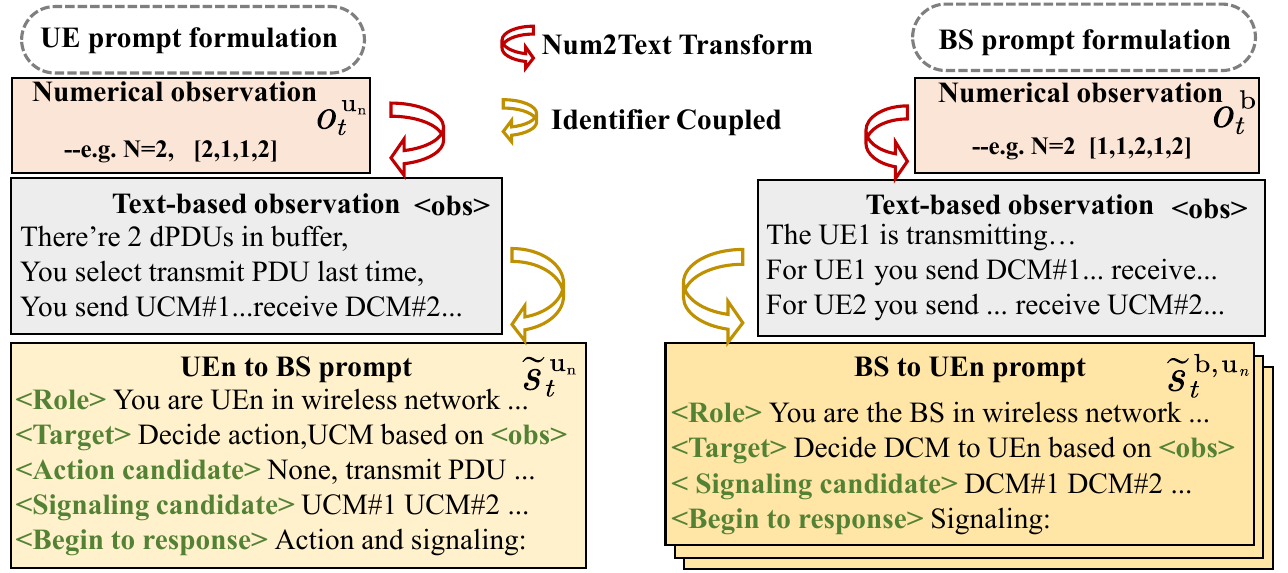}
    \caption{Numeric-Text transformation and prompt assembly.}
    \label{fig:prompt_formulation}
\end{figure}

\subsubsection{SIE-Enhanced Prompt}
\label{sec:id}
To avoid potential contextual ambiguity arising from concurrent prompts triggered by different UEs, we propose a hierarchical identifier architecture that systematically encodes network identity, contextual targets (e.g., service objectives, role tasks), and linguistic actions, and develop a comprehensive  
LLM-processable prompts as 
\begin{equation}
    \Tilde{s}\subscriptuet \triangleq \Tilde{o}\subscriptuet \oplus \text{Concat}\left(f_{\rm role}(n),f_{\rm tar}(g\subscriptuet),f_{\rm act}(\Tilde{\mathcal{A}}^{\UEn}) \right) ,
\end{equation}
\begin{equation}
    \Tilde{s}\subscriptbsuet{n} \triangleq \Tilde{o}\subscriptbst \oplus \text{Concat}\left(f_{\rm role}(n),f_{\rm tar}(g\subscriptbsuet{n}),f_{\rm act}(\Tilde{\mathcal{A}}^{\BS}) \right) ,
\end{equation}
where $g\subscriptuet,g\subscriptbsuet{n}$ represent primitive target (e.g., service objectives). Particularly, based on mapping functions (e.g., $f_{\rm role}$ and $f_{\rm tar}$, and $f_{\rm act}$), special tokens like \texttt{<Role>}, \texttt{<Target>} and \texttt{<Action candidate>} are explicitly inserted into prompts. The whole SIE process is illustrated in Fig. \ref{fig:prompt_formulation}.
\subsubsection{LLM-based Multi-Agent Policy}
At each time slot $t$, $\uepolicy(\Tilde{a}\subscriptuet|\Tilde{s}\subscriptuet)$ and $\bspolicy(\Tilde{a}\subscriptbsuet{n}|\Tilde{s}\subscriptbsuet{n})$ represent the policy of the $\UEn$ and BS, respectively, which are derived from the probability distribution of possible actions after feeding SIE prompts into the LLM. Taking $\UEn$ as an example, each action is represented as a sequence of tokens $\Tilde{a}\subscriptuet = \{w_0,\cdots,w_{\lvert \Tilde{a}\subscriptuet \rvert}\}$, $w\in\mathcal{W}$. The probability of a specific action is computed by
\begin{equation}
    \Pr(\Tilde{a}\subscriptuet \lvert \Tilde{s}\subscriptuet ) = \prod\nolimits_{j=0}^{\lvert \Tilde{a}\subscriptuet \rvert} \Pr(w_j \lvert \Tilde{s}\subscriptuet,w_{<j} )
    \label{eq:llm_generation}
    .
\end{equation}
Here, $\Pr(w_j \lvert \Tilde{s}\subscriptuet,w_{<j} )$ represents the probability of generating token $w_j$ given prompt $\Tilde{s}\subscriptuet$ and preceding tokens $w_{<j}$. We compute the log-probabilities from Eq. \eqref{eq:llm_generation} and then apply the softmax function to normalize these values. This yields the probability distribution over the action space $\mathcal{\Tilde{A}}^{\UEn}$ as
\begin{equation}
    \uepolicy(\Tilde{a}\subscriptuet \lvert \Tilde{s}\subscriptuet ) = 
    \frac{\exp\{\sum_{j=0}^{\lvert \Tilde{a}\subscriptuet \rvert} \log\Pr(w_j \lvert \Tilde{s}\subscriptuet,w_{<j})\}}
    {\sum_{\Tilde{a}\subscriptuet \in \mathcal{\Tilde{A}}^{\UEn}} \exp\{\sum_{j=0}^{\lvert \Tilde{a}\subscriptuet \rvert} \log\Pr(w_j \lvert \Tilde{s}\subscriptuet,w_{<j})\}
    }.
\end{equation}
Such policy formulation fully exploits the language model heads' prior and avoids potential ad-hoc mappings during text generation. Similarly, we can derive the BS policy $\bspolicy$. 

Furthermore, to merge fragmented agent policies, we introduce a unified policy fusion method as
\begin{equation}
    \llmpolicy = \frac{1}{Z}\sum_{e\in \mathcal{N} \cup \text{BS}} \underbrace{ {\rm exp}\left(-\varepsilon \cdot D_\text{KL}(\pi^{e}\lvert\lvert \pi^{\rm avg})\right)}_{\rm Consensus\ Factor} \times \pi^{e}(\Tilde{a}^{e} \lvert \Tilde{s}^{e})
    ,
\end{equation}
where $Z$ is the sum of consensus factors for normalization,
$\pi^{\rm avg} = \frac{1}{N+1}\sum_e \pi^{e}$ maintains temporal consistency. This operation resolves the fundamental tension between cross-agent coordination and individual policy expressiveness in a coordinated, efficient manner.
\subsubsection{PPO Alignment}
We replace the final layer of the initial LLM decoder block with a single-value head, multi-layer perceptron (MLP) to construct the critic network $V$, the input of which is $\Tilde{s}_t \triangleq (\Tilde{s}\subscriptuet[0],\dots,\Tilde{s}\subscriptuet,\Tilde{s}\subscriptbst)$. In each training epoch, the LLM interacts with the environment and stores transitions (i.e., Eq. \eqref{eq:rollout}) in a buffer. Once the buffer is gathered, we update the actor and critic via Eq. \eqref{eq:actor_loss} and Eq. \eqref{eq:critic_loss}, respectively.
\section{Simulation Results and Analysis}
\begin{table}[t]
    \caption{Simulation parameters}
    \centering
    \begin{tabular}{lcc}
    \toprule
    Parameter & Symbol & Value\\
    \midrule 
    Number of UEs & $N$ & $\{2,3,4,5\}$ \\
    Buffer size of each UE & $\lvert \mathcal{B}\rvert$ & $15$ \\
    Max. duration of episode (TTIs) & $T$ & $2
    4$ \\
    The arrive rate of dPDU & $p_a$ & $0.48$ \\
    Constant factor of reward & $\varrho$ & $1.5$ \\
    Vocabulary size of UCM & $U$ & $2$ \\
    Vocabulary size of DCM & $D$ & $3$ \\
    Discount factor & $\gamma$ & $0.99$ \\
    Transport block error rate & TBLR & $10^{-3}$ \\
    Clipping parameter of PPO & $\epsilon$ & $0.2$ \\
    \bottomrule
    \end{tabular}
    \label{tab:sim_paras}
\end{table}
\begin{figure}[t]
    \centering
    \includegraphics[width = 0.48\textwidth]{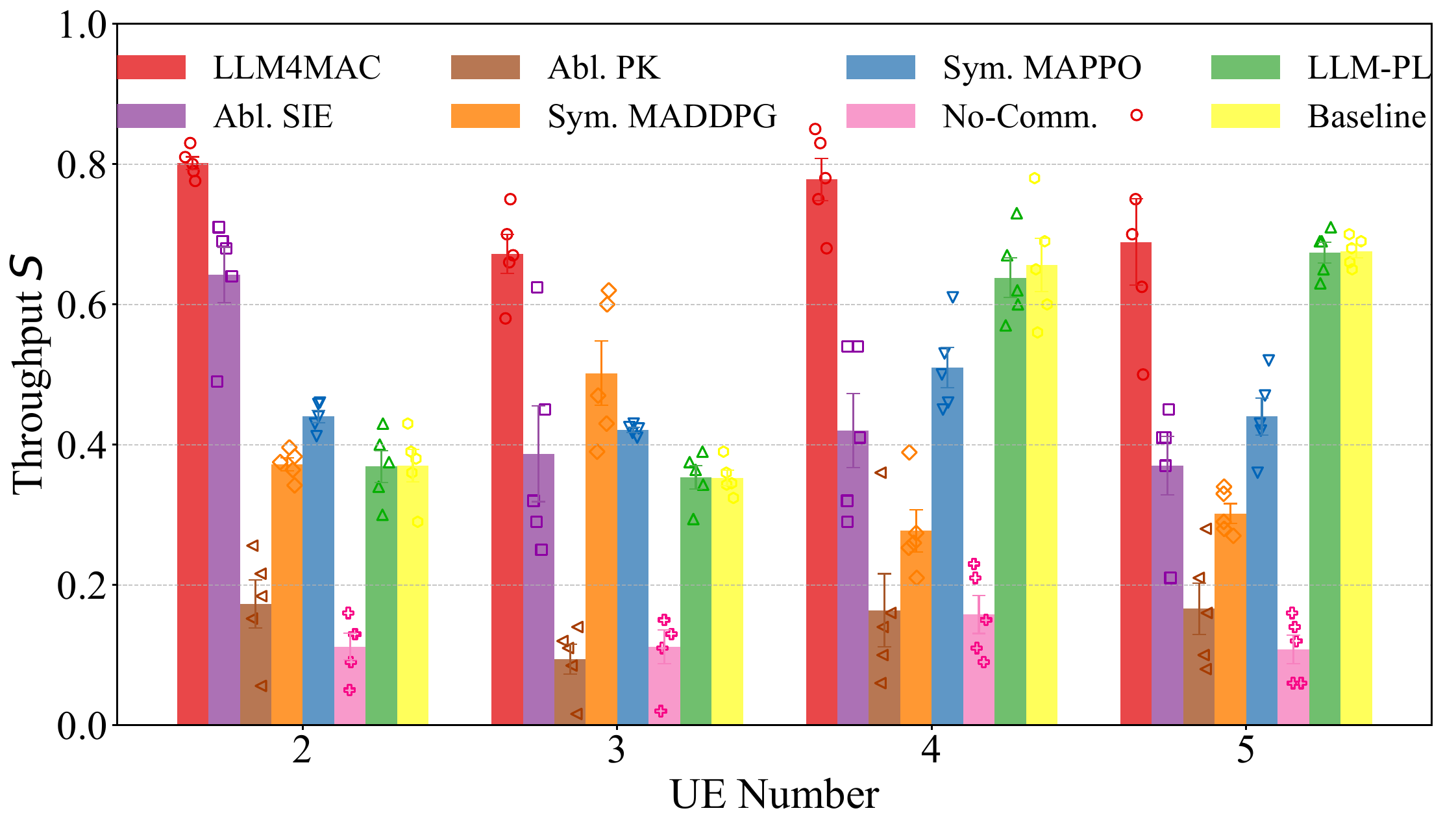}
    \vspace{-0.3cm}
    \caption{The comparisons of network throughput with other algorithms, where five independent experiments are performed for each case.}
    \label{fig:goodput}
\end{figure}
\begin{figure}[t]
    \centering
    \includegraphics[width = \linewidth]{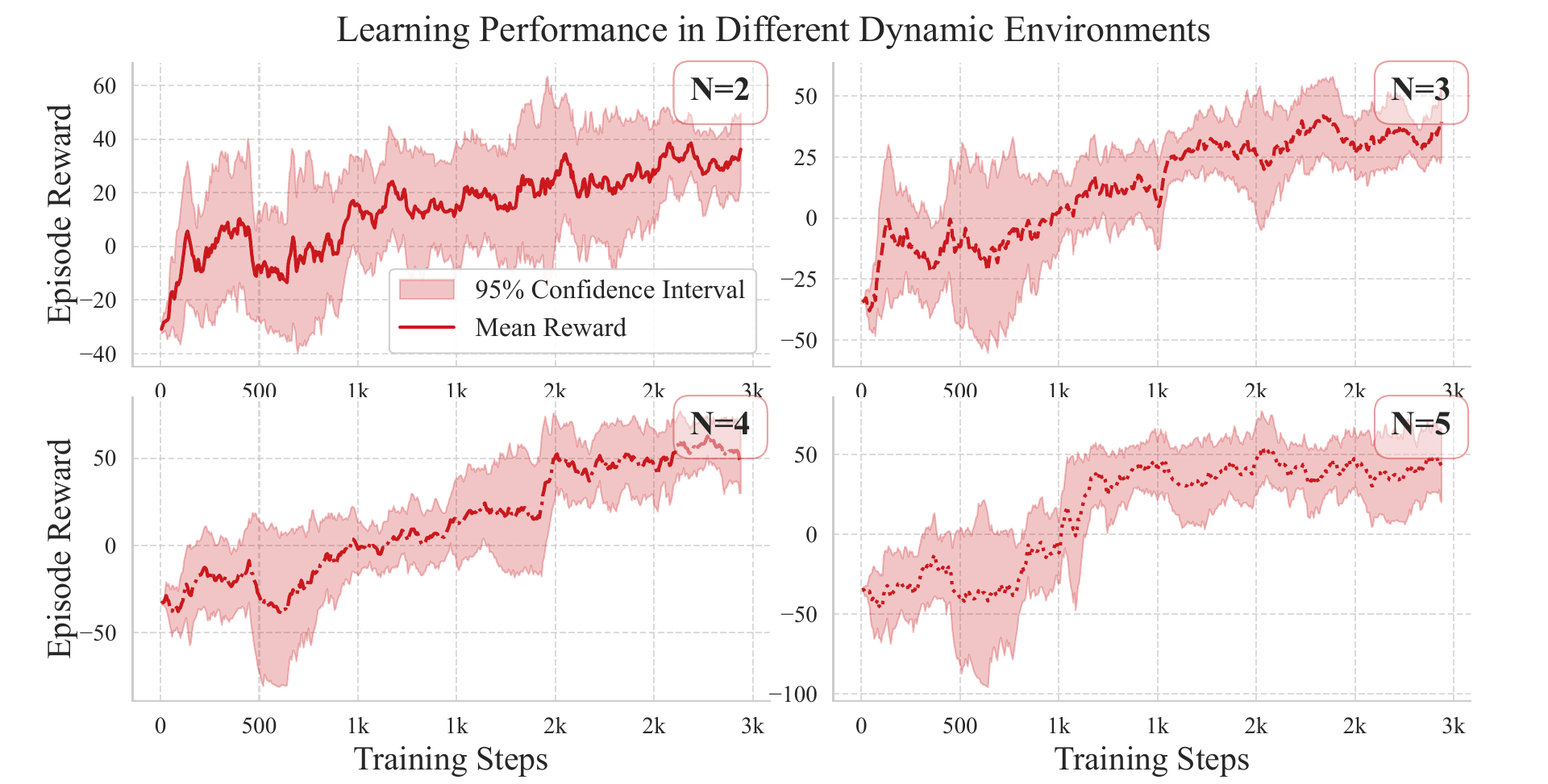}
    \vspace{-.6cm}
    \caption{Rewards v.s. training steps, where rewards are evaluated after each indicated training step. Experiments are conducted with different random seeds.}
    \label{fig:reward}
\end{figure}
In our simulation, a dynamic UDTS environment is constructed where the number of UEs $N$ fluctuates randomly within a predefined range $\{2,3,4,5\}$. The transmission buffer of each user starts empty and the episode length $T$ is $24$. We employ Flan-T5 base \cite{flant5} which is relatively compact and efficient for practical deployment, as the backbone of LLM4MAC, to learn both agent access and signaling policy. The system is trained for $3,000$ epochs, with periodic evaluations conducted in environments with fixed UE numbers to assess the learned protocol's performance. Simulation parameters are summarized in Tab. \ref{tab:sim_paras}. Meanwhile, we consider the following baselines.
\begin{itemize}
    \item \textbf{Baseline} - Human-crafted protocol same as Ref. \cite{MP-new}.
    \item \textbf{LLM-PT} - LLM prompt learning using a dataset collected via a human-crafted protocol \cite{MP-new}. 
    \item \textbf{Sym. MAPPO} - Symbolic MAPPO same as Ref. \cite{miuccioLearningGeneralizedWireless2022}.
    \item \textbf{Sym. MADDPG} - Symbolic MADDPG same as Ref. \cite{MP-scale}.
    \item \textbf{No-Comm.} - No signaling between UEs and BS.
    \item \textbf{Abl. PK} - Ablation for pretrained knowledge.
    \item \textbf{Abl. SIE} - Ablation for SIE mechanism.
\end{itemize}
To deliver as many dPDUs as possible within an episode, we evaluate the performance in terms of the network throughput in dPDUs/TTIs as
\begin{equation}
    S = \sum\nolimits_t \xi_{t,{\rm rec}}\diagup T.
\end{equation}
\begin{figure}[t]
    \centering
    \includegraphics[width = \linewidth]{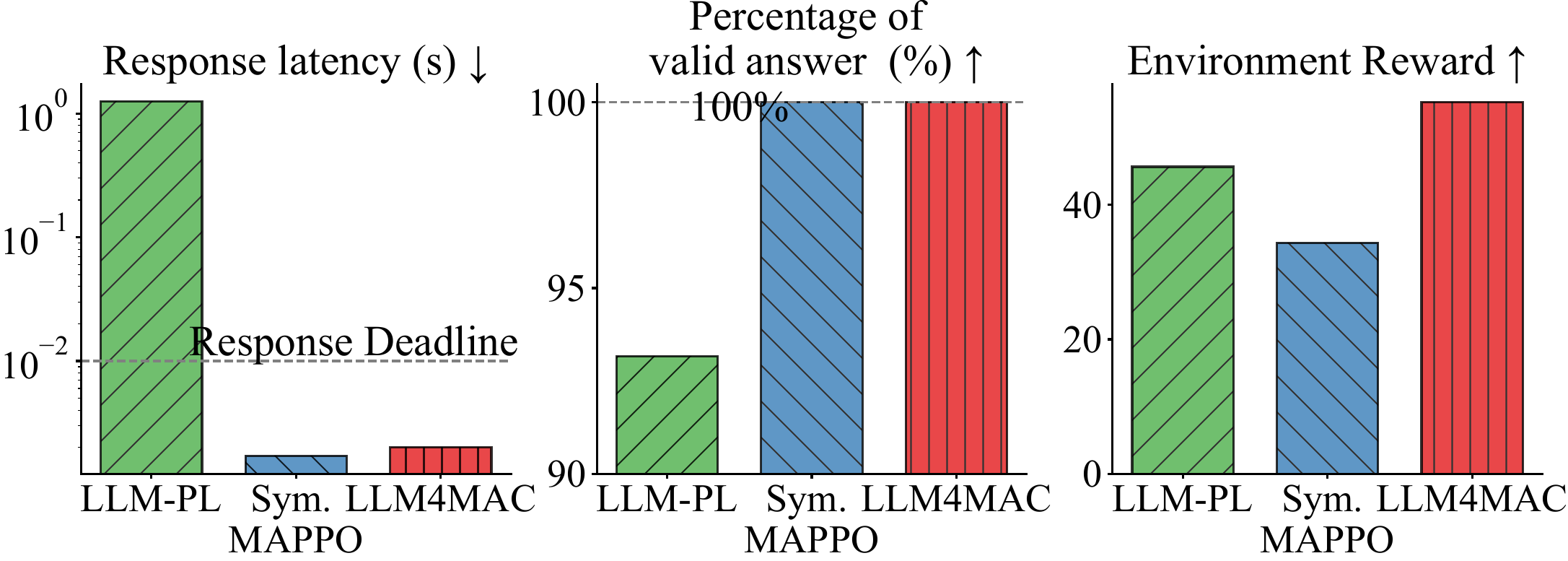}
    \vspace{-.6cm}
    \caption{The comparisons in terms of response latency (evaluated on Nvidia RTX 3090 GPU), valid answer percentage, and average environment reward.}
    \label{fig:error}
\end{figure}
Fig. \ref{fig:goodput} presents the comparative performance evaluation of LLM4MAC against several baseline approaches in terms of throughput $S$ across varying UE numbers. Notably, LLM4MAC consistently outperforms all other methods across all UE numbers. This demonstrates its effectiveness and generalizability in optimizing UDTS problems. Furthermore, as shown in Fig. \ref{fig:reward}, \texttt{LLM4MAC} converges unanimously and generalizes effectively in dynamic, heterogeneous agent environments. In contrast, other methods require re-training or multiple models.
In Fig. \ref{fig:goodput}, the performance degradation in \texttt{Abl. PK} and \texttt{Abl. SIE} validates the effectiveness of the SIE mechanism and pretraining. In particular, the impact of SIE becomes more pronounced as UE numbers increase, with throughput reductions of $22.5\%$ at $N=2$ and approximately $50\%$ at $N=5$. This highlights the contribution of identifiable roles towards fostering more effective interaction protocols. 
Furthermore, \texttt{LLM4MAC} consistently outperforms symbolic RL-based methods, such as \texttt{Sym. MAPPO} and \texttt{Sym. MADDPG}, achieving significant throughput gains across all UE numbers, whereas the latters struggle to adapt to dynamic environments. Meanwhile, \texttt{No-Comm.} yields the lowest performance, reaffirming the necessity of protocol-based communication for effective multi-agent coordination.
Due to the lack of RL-based exploration, \texttt{LLM-PL} experiences limited performance to break the upper bound of \texttt{Baseline}. 
Overall, these results underscore the advantages of \texttt{LLM4MAC} in facilitating cooperative multi-agent protocol emergence. 

Fig. \ref{fig:error} underscores the marked advantages of \texttt{LLM4MAC} across other core metrics — response latency, valid answer percentage, and average environment reward — under a fixed $N = 2$. In the left panel, \texttt{LLM-PL} leads to second-level latencies, far exceeding the duration of 5G-NR radio frames \cite{3GPP38321} and hindering real-time adaptation. By contrast, \texttt{LLM4MAC} attains microsecond-level responsiveness by merely using specific probabilities as in Eq. \eqref{eq:llm_generation}, instead of complete, contextual reasoning and  associated post-processing for generating responses in \texttt{LLM-PL} \cite{carta2024groundinglargelanguagemodels}. The middle panel further demonstrates that \texttt{LLM4MAC} consistently yields superior answer format accuracy, eliminating the physically invalid responses (``hallucinations”) that \texttt{LLM-PL} occasionally generates. Meanwhile, the right panel reveals that although  \texttt{Sym. MAPPO} avoids these pitfalls and meets rapid-response criteria, its overall performance lags behind, compounded by its requirement for frequent retraining in dynamic environments.
\section{Conclusion}
In this work, we have studied the emergence of efficient and generalizable MAC protocols in dynamic wireless networks. To achieve this, we have proposed an LLM4MAC framework, which integrates the exploration capabilities of RL with the generalization abilities of LLMs. The framework is built upon a semantics-generalized POMG and leverages a hierarchical, structured SIE architecture, along with a unified policy fusion mechanism, to facilitate multi-agent coordination. Simulation results confirm its superior performance compared to the baseline approaches. Future work will explore the generalizability of the scheme across a broader range of factors and further investigate its interpretability.


\end{document}